\documentclass[%
 aip,%
 apl,%
 amsmath,amssymb,
 floatfix,
 reprint%
]{revtex4-1}

\usepackage{graphicx}
\usepackage{float}
\usepackage{siunitx}
\usepackage{bm}
\newcommand{\Dp}{D_{\rm p}}
\newcommand{\Db}{D_{\rm b}}
\begin{document}


\title{Microscopic metallic air-bridge arrays for connecting quantum devices}

\author{Y. Jin}
\affiliation{Semiconductor Physics Group, Department of Physics, Cavendish Laboratory, University of Cambridge, Cambridge, CB3 0HE, UK}
\author{M. Moreno}
\thanks{These authors have contributed equally to this work.}
\affiliation{ Departamento de F\'{i}sica Aplicada, Universidad de Salamanca, Plaza de la Merced s/n, 37008 Salamanca, Spain}
\author{P.~M.~T. Vianez}
\thanks{These authors have contributed equally to this work.}
\author{W.~K. Tan}
\author{J.~P. Griffiths}
\affiliation{Semiconductor Physics Group, Department of Physics, Cavendish Laboratory, University of Cambridge, Cambridge, CB3 0HE, UK}
\author{I. Farrer}
\affiliation{Department of Electronic \& Electrical Engineering, University of Sheffield, Sheffield, S1 3JD, UK}
\author{D.~A. Ritchie}
\author{C.~J.~B. Ford}
\thanks{Author to whom correspondence should be addressed: cjbf@cam.ac.uk}
\affiliation{Semiconductor Physics Group, Department of Physics, Cavendish Laboratory, University of Cambridge, Cambridge, CB3 0HE, UK}


%

\date{\today}

\begin{abstract}
We present a single-exposure fabrication technique for a very large array of microscopic air-bridges using a tri-layer resist process with electron-beam lithography. The technique is capable of forming air-bridges with strong metal-metal or metal-substrate connections. This was demonstrated by its application in an electron tunnelling device consisting of 400 identical surface gates for defining quantum wires, where the air-bridges are used as suspended connections for the surface gates. This technique enables us to create a large array of uniform one-dimensional channels that are open at both ends. In this article, we outline the details of the fabrication process, together with a study and the solution of the challenges present in the development of the technique, which includes the use of water-IPA (isopropyl alcohol) developer, calibration of resist thickness and numerical simulation of the development.

\end{abstract}

\maketitle

The interconnection of conducting layers is key to the performance of printed or integrated circuits. When fabricating ultra-small specialised research devices, however, the process for depositing and patterning an insulating layer to keep regions apart, or to space gates away from the surface in places, is complex and often affects operation. A straightforward and reliable method for bridging between regions is therefore highly desirable and can make possible much more complicated device architectures for physics research. This is particularly needed in areas such as quantum computing, where interconnecting qubits can often prove challenging.

Normal\cite{Ford1989-1998} and cross-linked\cite{Zailer1996-1887,Kataoka2002} polymer resist (PMMA) has been used as a patterned insulator under metal bridges for studying quantum ring structures and antidots. Air-bridge structures have also been used in quantum-dot interference devices\cite{Yacoby1994,Ji2003,Gurman2016}, where the bridge played a crucial role in connecting a central gate while leaving the interference path undisturbed, and superconducting microwave circuits based on coplanar waveguides, where the use of bridges prevents the propagation of parasitic modes, hence reducing the amount of loss and decoherence. Here, however, even though workable techniques have already been proposed in the literature, their significant degree of complexity means that in practice they are still rarely used.\cite{holman_microwave_2020,chen_fabrication_2014,lankwarden_development_2012} Various methods have been employed for the fabrication of these bridges: \textcite{Li2001}, for example, demonstrated a process with nanoimprint lithography (NIL) for monolithic microwave integrated circuits with air-bridges. While NIL simplifies repeated fabrication, its complexity is unsuitable for rapid iteration of research prototypes. Instead, resist exposure is preferable. Photo-resist can be partially cross-linked and later removed to form an air gap below the bridge.\cite{Jie2012} One drawback of cross-linking, however, is that it is susceptible to pattern distortion due to swelling of the resist,\cite{Brewer1980} hence making it unsuitable for dense sub-micron patterns. 

An alternative approach is variable exposure: the electron-beam penetration depth into the resist can be controlled by varying the acceleration voltage and dose on a single layer of PMMA resist in a one-stage exposure,\cite{Girgis2006} and polyimide and double-layer PMMA can be combined in a two-stage technique.\cite{Khalid2012262} In the former method, varying the acceleration voltage has the undesirable consequence of changing the focus and alignment and the low acceleration voltage limits the electron-beam resolution. The latter method does not suffer from these drawbacks but does involve the use of two lithography stages. 

In this letter, we present a process with the combined advantages of the single-stage exposure and multiple-resist methods, in order to fabricate large numbers of fine-feature air-bridges with very high yield. We have optimized the process by using a water/IPA (isopropyl alcohol) mixture to develop the PMMA. This minimizes residual resist and gives good exposure contrast. We show results from a set of 1D wires defined using an array of gates linked by $\sim$\, 400 air-bridges. We note however that we have regularly used this technique to fabricate devices with up to $\sim$\,6000 bridges. This technique can also be used to fabricate bridges up to at least 5\,$\mu$m in length and therefore should be useful in a wide range of nanodevices made for research purposes.

\begin{figure*}
	\includegraphics[width=\textwidth]{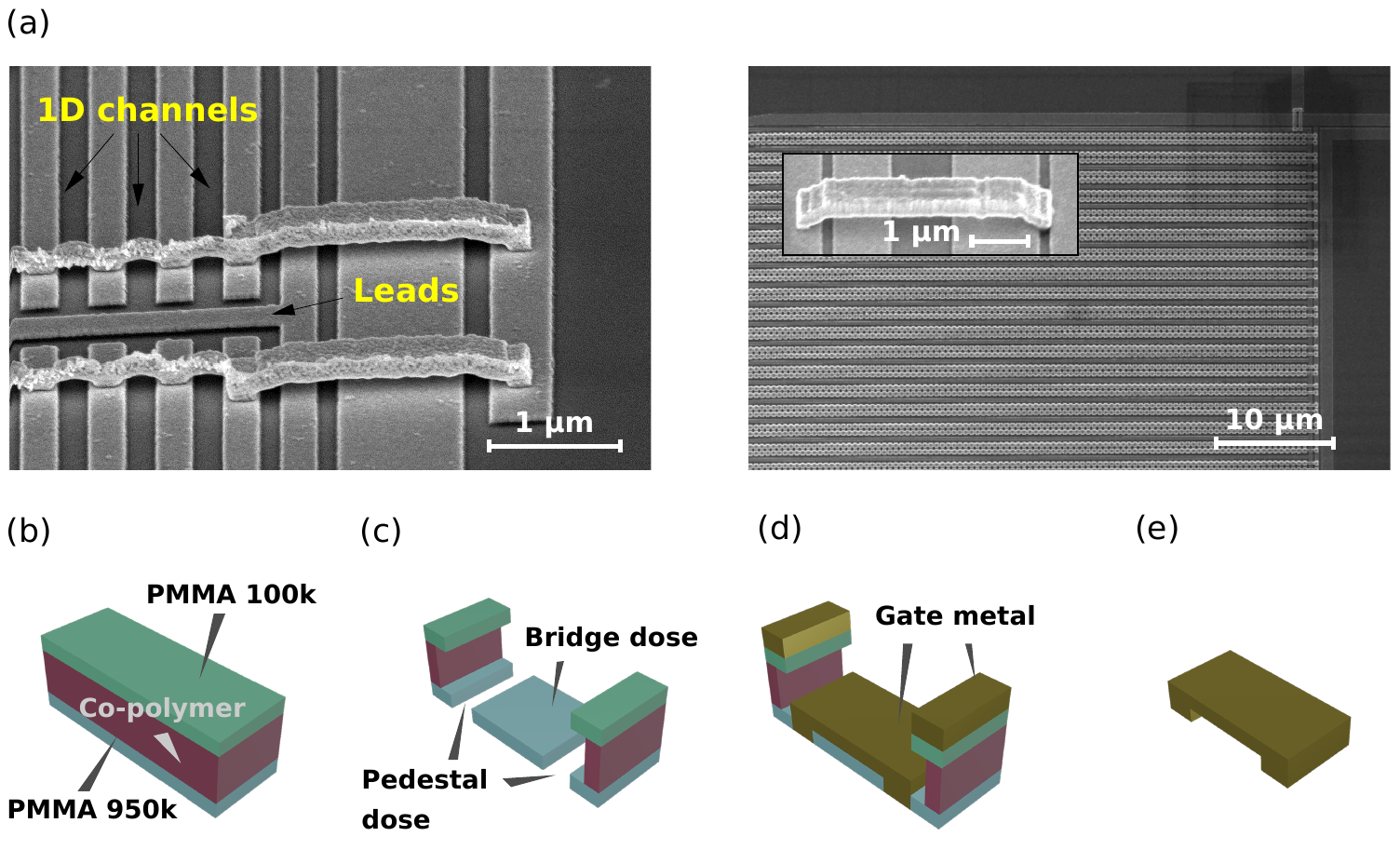}
\caption{(a) SEM images of a 1D tunnelling device. 1D channels are formed beneath the narrow regions between two depletion gates, with current being injected via wide, 2D leads under the gate labelled `leads'. The tunnel device contains multiple columns of identical depletion gates that are inter-connected by air-bridges without closing the channels, as well as crossing over other device structures. Inset: SEM micrograph of a bridge 5\,$\mu$m long [accidental misalignment put the right-hand pedestal in a gap]. (b)--(e) Air-bridge fabrication by our three-layer PMMA single-exposure process: (b) Triple-layer resist spin-coated on sample surface. (c) Selective exposure by  electron beam and development. Regions exposed to pedestal dose are completely cleared after development while regions exposed to bridge dose have the 950k layer intact. (d) Gate metallization by thermal evaporation. (e) Lift-off leaves just the air-bridge pattern. \label{3layer_steps}}
\end{figure*}

The development of our air-bridge technique was motivated by the need to fabricate arrays of 1D channels in order to study the exotic properties of electron-electron interactions, specifically regarding non-Fermi-liquid behaviour and the Luttinger liquid model.\cite{Moreno2016,Tsyplyatyev2016,Tsyplyatyev2015,Jin2019,Vianez_2021} Figure 1a demonstrates the geometry of the array under scanning electron microscopy (SEM). The substrate is a GaAs/AlGaAs heterostructure that contains two parallel quantum wells separated by a \SI{14}{\nano\meter}-thick tunnel barrier, which allows electron tunnelling. A 1D electron channel is formed underneath the narrow region between each pair of metallic gates when they are negatively biased. Each device consists of a large number of identical channels organized into multiple sets of parallel wires to enhance the tunnelling current. We require the gates to be electrically connected while keeping the potential in the 1D channel as uniform as possible along its whole length. Consequently, bridges are necessary. An air gap, instead of a solid dielectric, provides minimal capacitive coupling between the bridge and the 2D electron gas (2DEG) in the quantum well underneath, for a given gap height. Approximating the structure to a parallel-plate capacitor, the capacitance is given by $C=\varepsilon_0 A/(d_{\rm sp}/\varepsilon_{\rm sp} + d_{\rm sub}/\varepsilon_{\rm sub})$, where $A$ is the area of the bridge, $\varepsilon_{\rm sp}, \varepsilon_{\rm sub}$ are the dielectric constants of the spacer below the bridge and of the substrate above the 2DEG, respectively, and $d_{\rm sp}, d_{\rm sub}$ are their thicknesses. Therefore, $C$ is minimized by minimizing $\varepsilon_{\rm sp}$ (using an air gap), and maximizing $d_{\rm sp}$.

Figure 1b-1e outlines the steps of our multilayer-resist/single-exposure air-bridge process. It begins with the spin coating of the sample with three different resist layers, firstly PMMA with molecular weight 950k, then MMA(8.5)MAA copolymer and finally 100k PMMA (Figure 1b). Next, the sample is patterned by electron-beam lithography (EBL) using two well calibrated doses $\Dp$ and $\Db$, referred to as the pedestal dose and the bridge dose, respectively. $\Dp$ is capable of fully exposing all three resist layers, while $\Db$ is only able to expose the top two and leaves the bottom layer unaffected, because 950k PMMA has much lower sensitivity than the other resists. The resist profile after development is shown in Figure 1c, where areas exposed by $\Dp$ have been completely cleared and those exposed by $\Db$ are still covered in 950k PMMA. The copolymer is far more sensitive than the PMMA and so the middle layer is undercut relative to the top layer, which aids with the removal of residual metal during lift-off. An air-bridge structure is formed when gate metal is evaporated on top (Figure 1d) and remains on the sample after the resists are stripped off (lift-off, Figure 1e). Metal over the $\Dp$ region is in direct contact with the substrate, and is referred to as the pedestal. `Bridge` regions exposed at $\Db$ are covered in metal separated from the substrate by an air gap but anchored to the substrate via pedestals. The thickness of the bottom layer of resist after development corresponds to the height of the air gap below the bridge, which in our samples is approximately $\sim$\,100\,nm.

In order to achieve a reliable process, it is necessary to precisely control the thickness and uniformity of the resists. Using ellipsometry, we calibrated the thickness as a function of spinner rotation speed for each type of resist using test wafers. Additionally, after each layer was spin-coated on the actual sample, an ellipsometry measurement was performed to confirm that the resist was within $\pm$\SI{10}{\nano\meter} of the target thickness. For the reported sample, we applied \SI{133}{\nano\meter} of 950k, \SI{297}{\nano\meter} of copolymer and \SI{128}{\nano\meter} of 100k resists using 60-second spins at 5700, 4500 and 6000 RPM, respectively. The target thicknesses were 130, 300 and \SI{130}{\nano\meter}. We note, however, that the thickness can be changed and working samples have been obtained within $\pm$ 20\,\% of these values after adjusting for the e-beam dose. The dilution ratios of the resist solutions were: 4\% (w/w) 950k PMMA in anisole diluted with methyl isobutyl ketone (MIBK) in 2:1 volume ratio, 9\% (w/w) copolymer in ethyl lactate and 100k PMMA (undiluted). The base doses received by the resist for the exposure of the pedestals and bridges were 880 and 600\,$\mu$\textrm{C}/\textrm{cm}$^2$, respectively. Prior to the resists being applied, the sample was baked on a \SI{150}{\celsius} hotplate for 10 minutes to eliminate moisture. After each layer was applied, the sample was also further baked on a \SI{110}{\celsius} hotplate for 10 minutes to dry off solvents in the resist. Finally, we note that when spinning, particularly for the 100k resist, significantly better uniformity was obtained when using a metal as opposed to a PTFE stage.

\begin{figure} 
	\includegraphics[width=\columnwidth]{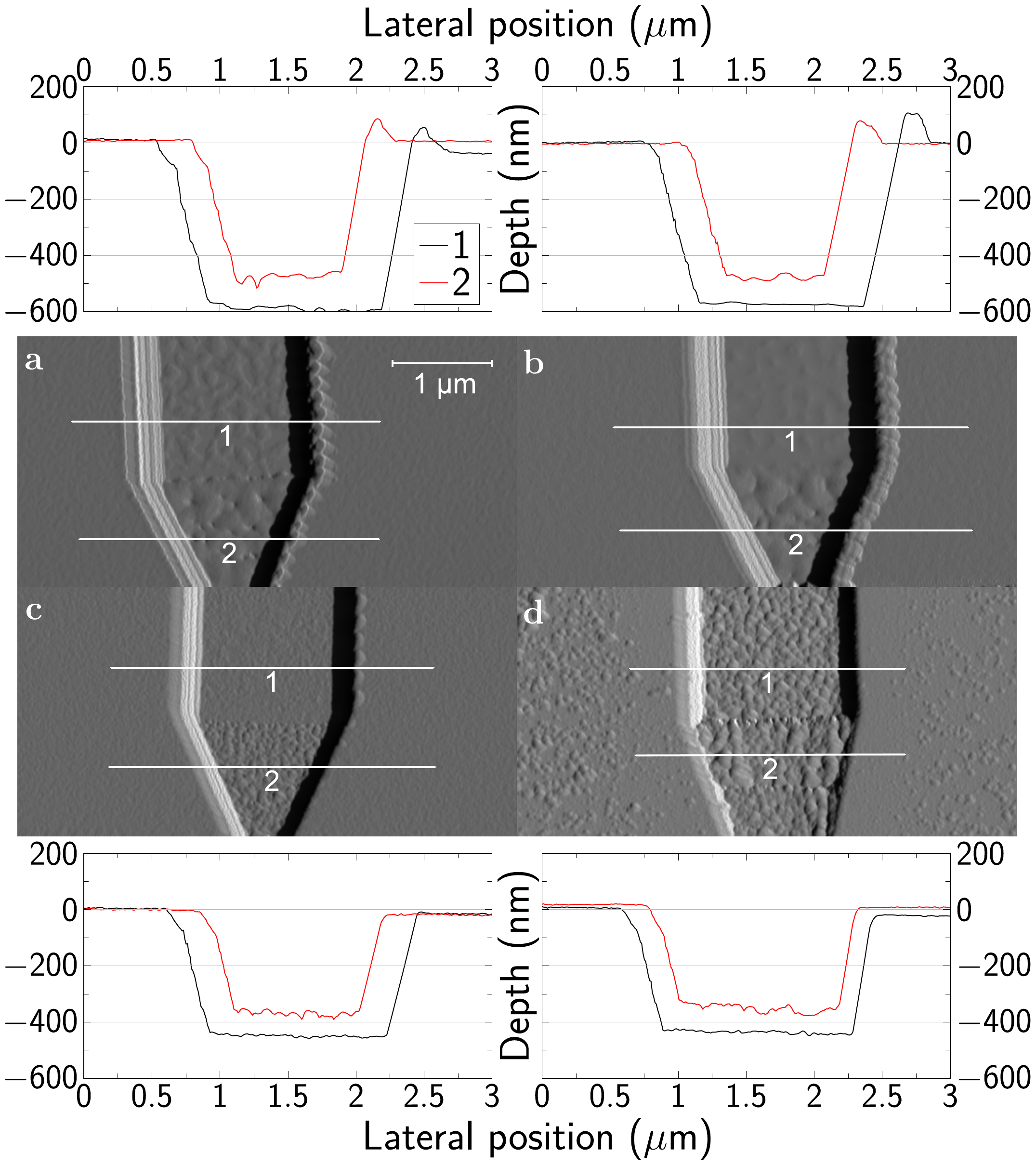}
	\caption{Differentiated AFM images of resist patterns where the upper part of the pattern (line 1, black in accompanying cross-sections) received the pedestal dose and the lower, tapered part (line 2, red) received the bridge dose. Various combinations of resist and development were compared: (a) Three layers of resist (100k, copolymer, 950k) developed at room temperature ($\sim$ \SI[separate-uncertainty = true]{21}{\celsius}) for 30 seconds in 3:1 IPA/MIBK. (b) The same sample as in (a) after 15 seconds of plasma ashing at \SI{50}{\watt}. (c, d) Two layers of resist (copolymer, 950k) developed at room temperature for 5 seconds in (c) 3:7 water-IPA and (d) 3:1:1.5\% IPA/MIBK/MEK (methyl ethyl ketone).
	\label{AFM}}
\end{figure}

The most significant challenge of the air-bridge process is achieving good adhesion between the pedestals and the underlying material. For our device this meant the metal-to-metal contact between the pedestals and the 1D channel gates. The standard development technique with $3:1$ MIBK:IPA developer was found to be unreliable as the resulting air-bridges often broke away from the sample during lift-off. Atomic force microscopy (AFM) showed the cause of this type of failure to be trace amounts of residual resist after development. Figure 2 shows a comparison of AFM scans of the same exposure pattern on 950k PMMA treated by different developers. The EBL doses and development times used in these results were such that the substrate was expected to be fully exposed after development. The images demonstrate that the choice of developer has substantial impact on the surface roughness of the developed region. As is shown in Figure 2b, the surface roughness can be reduced by RF plasma ashing, suggesting these are resist residues. Since plasma ashing attacks both the residue and the unexposed resist, it may also cause damage to the ultra-fine resist patterns. The safer option is therefore to adopt the developer that leaves the least amount of resist residue after development. The use of water/IPA mixture to develop PMMA was studied by Yasin.\cite{Yasin2001, Yasin2002745} Owing to the high sensitivity of the water/IPA developer at room temperature, we conducted the development at \SI[separate-uncertainty = true]{5.(5)}{\celsius} with the use of temperature-controlled water bath, in order to limit the rate of development and increase the tolerance to timing error in the process. The lower temperature also results in higher contrast.\cite{Yasin2002745} The samples were immersed in pre-cooled beakers of water/IPA mixture with manual agitation. No rinse was used at the end of development, as immersing the sample in pure water or IPA after development can increase the development rate and was also found to deposit precipitates on the sample surface. The best result was produced when the samples were removed immediately from the developer and dried with nitrogen gas.

\begin{figure}
 \centering
	\includegraphics[width=\columnwidth]{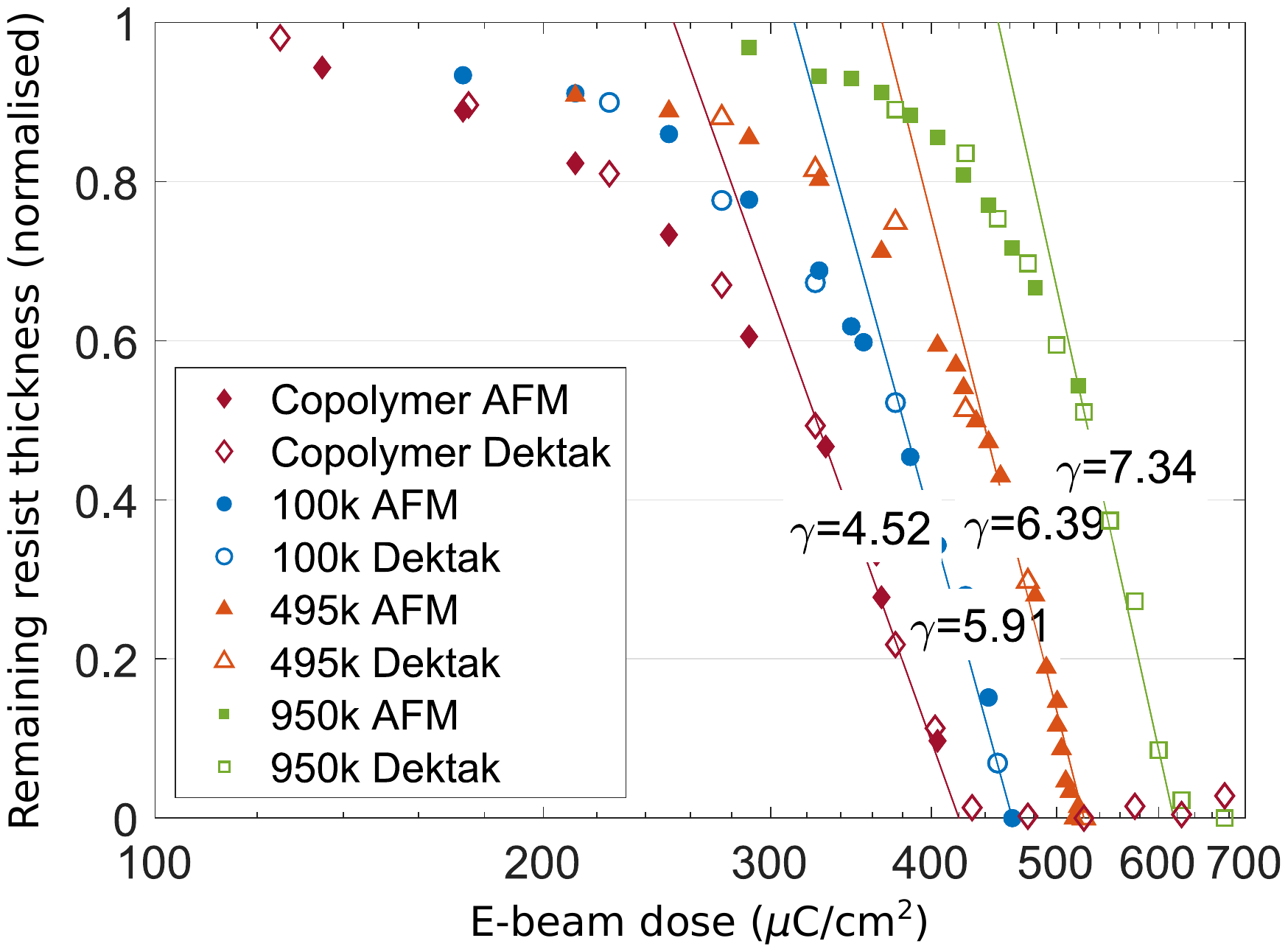}
\caption{Contrast curves of different EBL resists based on development depth after 30 seconds of immersion at \SI{5}{\celsius} in 3:7 water/IPA. Each data series is normalized to the original thickness of the corresponding type of resist (here, \SI{146}{\nano\meter} for 950k, \SI{313}{\nano\meter} for copolymer and \SI{105}{\nano\meter} for 100k resists). The contrast $\gamma$ refers to the gradient of the linear region of the curve. \label{logcurve}}
\end{figure} 

In order to calibrate $\Dp$ and $\Db$ for the triple-layer process, we determined the sensitivity curves of the three types of EBL resist by measuring the depths of developed resist as functions of electron-beam dose. Measurements were made on \SI[product-units = power]{200 x 300}{\micro\metre} rectangular test patterns, as well as fine gratings with 2\,$\mu$m width and 2\,$\mu$m separations. Both types of patterns were exposed with \SI{100}{\kilo\volt} acceleration voltage on a Vistec VB6 system over a range of doses. The development depth was measured both by a Dektak surface profiler and an AFM, with the former method applied to the rectangular patterns and the latter to the fine gratings. Figure 3 shows the normalized development depth of different types of resist as functions of EBL dose. When dose is plotted on logarithmic scale, the negative of the gradient of the linear part of the curve is the contrast $\gamma$.\cite{Brewer1980} We note that the contrast measured in our calibration is similar to the values reported by Rooks.\cite{Rooks2002} A polynomial fit to each of these curves was used to estimate the rate of development at any arbitrary dose. To give insight into the development process, we have developed a numerical model of the process with the electron-beam dose and development time as input. Our calculation assumes: 1.\,the development is a time-limited process, meaning development depth is always proportional to time; 2.\,the development has a uniform rate as a function of dose, and is estimated from the contrast curve; 3.\,the direction of development at each point is normal to the surface there. Figure 4 shows the result of this numerical calculation by displaying the evolution of the resist profile in 5-second increments. The calculation gives a similar hump of copolymer as seen in an under-developed sample shown in the SEM image in the inset.

Calculations performed using the average development rates from the measurements in Figure 3 imply that the optimum development time should be around \SI{35}{\second}. However, we found empirically that \SIrange{60}{70}{\second} was required to fully develop the combined layers, with poor metal adhesion for development times at or below \SI{55}{\second}. This discrepancy is likely caused by variations in the rate as development progresses: solute builds up in the developer, slowing down the dissolution. Hence different structures or depths may require different development times. Practically however, this can be managed by dividing samples into multiple development batches and using an iterative scheme to home in on the optimum time for a particular sample or type of device.

After development, we evaporated approximately 110--130\,nm (i.e. roughly equal to the thickness of the base resist) of gold at a rate of $\sim\,$0.3\,nm/s. The samples were then left overnight in a bottle of acetone, followed by a water bath at \SI{45}{\celsius} for approximately 90min before final lift-off. Use of ultrasonication is not recommended at this stage as this was found to often result in damage to the bridges.

\begin{figure}
    \centering
	\includegraphics[width=\columnwidth]{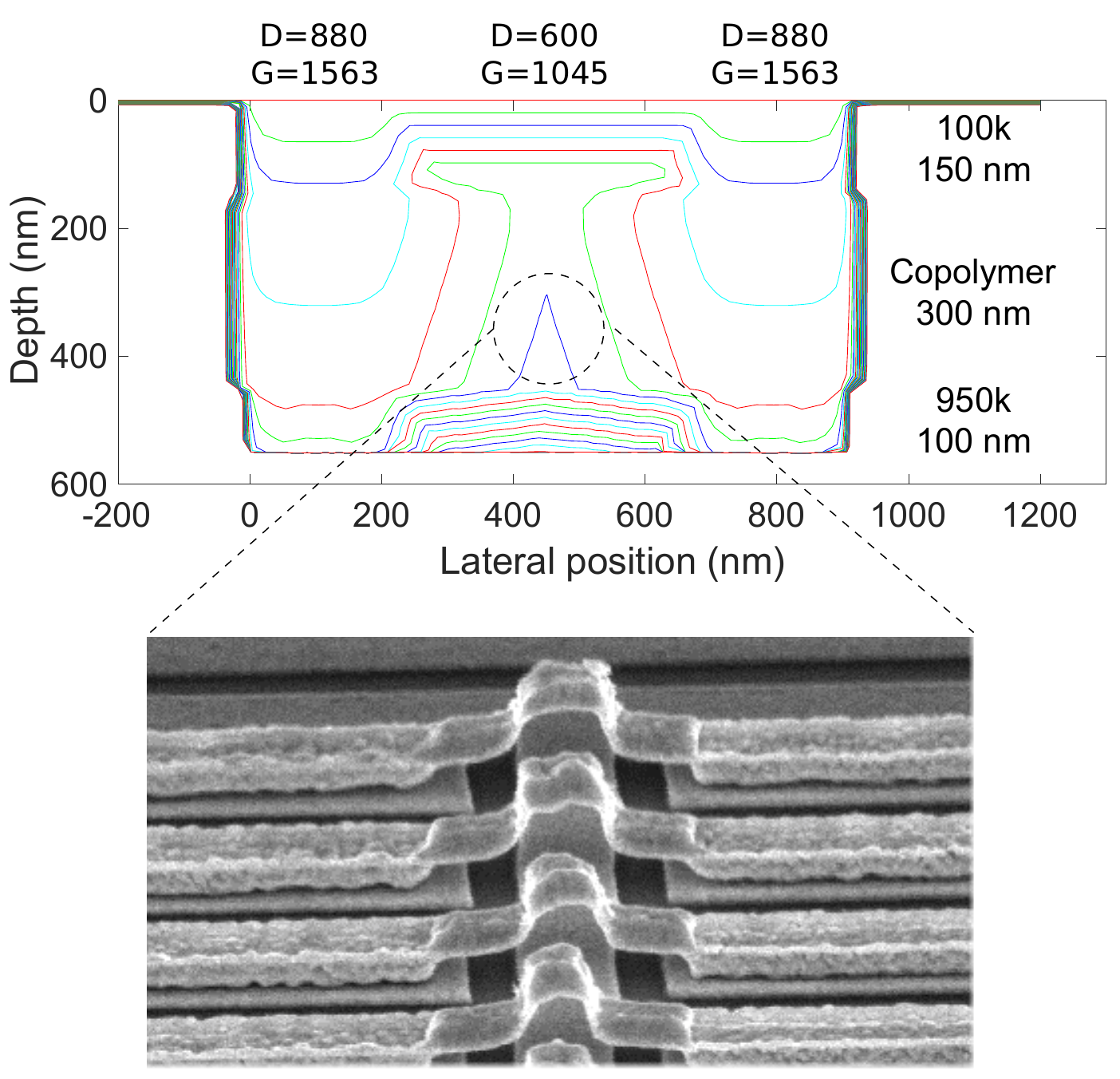}
\caption{Results of a numerical model estimating the evolution of resist profile as a function of electron-beam dose and time. Each contour differs from its closest neighbour by 5 seconds in development time. The figures at the top of each region are the average EBL dose $D$ required there, and the actual dose $G$ given by the EBL machine after correcting for proximity-effect. The same spreading parameters were used in both the model calculation and the proximity correction. Inset: SEM micrograph of an under-developed sample showing a similar hump of copolymer as predicted by our model.
\label{dev_model}}
\end{figure}

We checked the integrity of the air-bridge arrays by inspecting the sample under an optical microscope. Except for sacrificial trial samples, we refrained from analysing any experimental device under SEM, so as to avoid potential contamination by electron-beam-induced deposition. Although individual elements of the array cannot be clearly resolved under an optical microscope, the large number of repeating units produces a uniform and iridescent appearance of the entire structure, which can be very easily resolved on top of the surface-gate metal. In practice, optical inspection can therefore easily reveal defects in either of the two EBL layers, with the most common modes of failure being incomplete lift-off after the base layer metallisation, and poor adhesion between the air-bridge pedestals and the underlying base-layer metal. Both types of failure lead to defects that distort the uniform appearance of the array and are easily visible under optical microscopy. After optical inspection, samples were tested further for electrical continuity between contacts. The air-bridge-connected wire gates did not short to other nearby control gates and controlled the measured conductance in the ways expected.

\begin{figure*}
	\includegraphics[width=\textwidth]{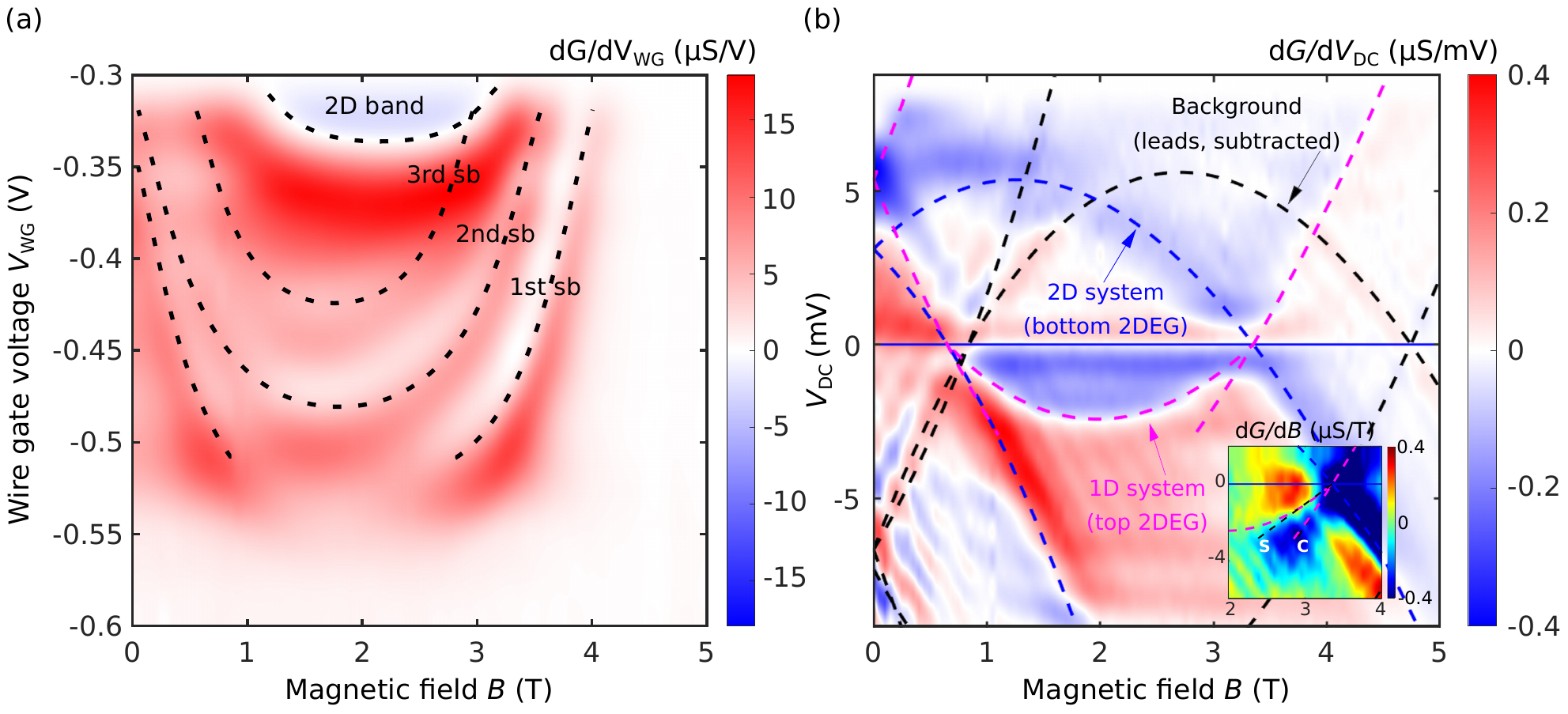}
	\caption{(a) Derivative of the equilibrium tunneling conductance $G$ of the resonance device with respect to the 1D wire-gate voltage $V_{\mathrm{WG}}$ \textit{vs} magnetic field $B$ and $V_{\mathrm{WG}}$, at $T=\SI{0.3}{\kelvin}$ and normalised along $B$. The black dashed curves mark the 2D band and the 1D subbands. Three 1D subbands can be resolved, with the last subband cutting off at $V_{\mathrm{WG}}\sim\SI{-0.55}{\volt}$. (b) Tunnelling differential $\textrm{d}G/\textrm{d}V_{\textrm{DC}}$ \textit{vs} $B$ and the DC bias $V_{\mathrm{DC}}$ between the two quantum wells, with $V_{\mathrm{WG}}$ held at $\SI{-0.515}{\volt}$. The parabolae indicate the positions of the calculated dispersions of the 1D and 2D systems. At this wire-gate voltage, only a single 1D subband is observed. The magenta/blue parabolae mark the observed dispersions resulting from 1D-2D tunnelling in the wire regions. The black parabolae, on the other hand, correspond to the dispersions of the background `parasitic' tunnelling region, arising from 2D-2D tunnelling in the leads. This signal can be separately mapped and subtracted, as shown. Inset: separation of spin and charge modes (dashed black `S' and `C' lines) at low-bias.}
	\label{subbands}
\end{figure*}

Finally, we present some typical measurements of one of our air-bridge devices. Figure 5a shows an intensity map of the equilibrium tunnelling conductance (i.e.{} at $V_{\textrm{DC}}=0$) as a function of $B$ and $V_{\mathrm{WG}}$ at temperature $T=0.3$\,K. The dashed lines highlight the positions of the local maxima of the tunnelling conductance $G=\textrm{d}I/\textrm{d}V_{\textrm{DC}}$. Figure 5b shows $\textrm{d}G/\textrm{d}V_{\textrm{DC}}$ \textit{vs} $B$ and $V_{\textrm{DC}}$, for $V_{\mathrm{WG}}=\SI{-0.515}{\volt}$. In our devices, while the bottom 2DEG always remains 2D in nature, the top 2DEG has regions that are confined (1D, under the wires) or unconfined (2D, elsewhere). Electrons must tunnel into empty states, so the maximum conductance is observed when the Fermi energy and wave vector of one (2D) system track the dispersion of the other (1D) system, revealing the dispersions of the 1D electron subbands, which, as expected, are essentially parabolic. Detailed fitting of these dispersions reveals their modification by strong electron-electron interactions in the 1D wires, including separate spin and charge modes (see inset in Fig.\,5b). 

Careful fitting of the 1D parabola in Fig.\,5b reveals that the 1D parabola below the $B$ axis does not extend smoothly above the axis. Instead, the dispersion above the axis at high $B$ extends down and matches the charge line (C in the inset). We can also see its equivalent coming from $B<0$ at low fields around 6\,mV and use this as a constraint to fit a set of identical parabolae. We interpret this parabola as the dispersion of a Fermi sea of charge excitations, independent of the spin excitations described by the original parabola.\cite{Vianez_2021} This is consistent with one of the theories of nonlinear 1D systems and provide evidence for a remarkably simple way of visualising the effect of strong correlations in low-dimensional systems.

The use of air-bridges to join the wire gates together was crucial when it came to the very short ($<3\,\mu$m) 1D channels, where a surface connection at one end of the wires would have caused too much depletion of parts of each wire and hence great non-uniformity. Since $G$ is summed over the $\sim$\,400 1D channels in a single device, the fact that that we are able to resolve the 1D subband structure clearly (Fig.\,5a) demonstrates the high degree of uniformity of the wire-gate array, and that the air-bridge structure connecting the gate array performs reliably. If this were not the case, for instance due to a break in the chain of bridges, then the whole of the array beyond that point would be disconnected, effectively staying at zero gate voltage, and therefore remaining 2D in nature, with an electron density similar to that of the leads. We are, however, capable of also separately tuning the lead density via its own surface gate, which means that if any significant portion of the array were disconnected, this would become very quickly apparent from the presence of extra high-field parabolae, which we do not observe.

In conclusion, we have demonstrated a process that is capable of reliably fabricating large arrays of air-bridges in a single step of EBL exposure. The essential steps to the process are: 1.\ Careful control of the thickness of deposited resist; 2.\ Accurate calibration of the dose curve; 3.\ Use of a water/IPA developer to improve the adhesion of the air-bridge pedestals. The process is suitable for fast iteration of prototype or research devices and can be generalised to other substrate materials and metals.

Acknowledgements: The authors would like to thank Yunchul Chung (Pusan National University, South Korea) for advice on air-bridge fabrication. This work was supported by the UK EPSRC [Grant Nos.\ EP/J01690X/1 and EP/J016888/1]. P.M.T.V.\ acknowledges financial support from EPSRC International Doctoral Scholars studentship via grant number EP/N509620/1.

\section*{Data Availability}
The data and modelling code that support this work are available at the University of Cambridge data repository (doi.org/10.17863/CAM.66595).

\bibliography{citations}

\end{document}